\documentclass[reprint,nofootinbib,superscriptaddress,amsmath,amssymb,aps,prd]{revtex4-1}
\usepackage{amsmath,mathrsfs,amsbsy,color,graphicx,bm,amsthm,amsfonts}
\usepackage{bbm}
\usepackage{times}
\usepackage{units}
\usepackage{dcolumn}
\usepackage{graphicx}
\usepackage{epsfig}
\usepackage{epstopdf}
\usepackage[colorlinks,linkcolor=blue,anchorcolor=green,citecolor=blue,CJKbookmarks=True]{hyperref}
\DeclareMathSymbol{\shortminus}{\mathbin}{AMSa}{"39}
\usepackage{mathrsfs}
\usepackage{braket}
\usepackage{amssymb}
\usepackage{txfonts}
\usepackage{float}
\usepackage{enumitem}
\usepackage{multirow}

\begin{document}
	\title{ Generation of quantum entanglement in superposed diamond spacetime}

	\author{Xiaofang Liu}
	\affiliation{Department of Physics, Key Laboratory of Low Dimensional Quantum Structures and Quantum Control of Ministry of Education, and Synergetic Innovation Center for Quantum Effects and Applications, Hunan Normal
		University, Changsha, Hunan 410081, P. R. China}
	
	\author{Changjing Zeng}
	\affiliation{Department of Physics, Key Laboratory of Low Dimensional Quantum Structures and Quantum Control of Ministry of Education, and Synergetic Innovation Center for Quantum Effects and Applications, Hunan Normal
		University, Changsha, Hunan 410081, P. R. China}
	
	\author{Jieci Wang}
	\email{jcwang@hunnu.edu.cn}\affiliation{Department of Physics, Key Laboratory of Low Dimensional Quantum Structures and Quantum Control of Ministry of Education, and Synergetic Innovation Center for Quantum Effects and Applications, Hunan Normal
		University, Changsha, Hunan 410081, P. R. China}
	
	\begin{abstract}
A comprehensive study integrating the microscopic structure of spacetime and the principle of quantum superposition is capable of offering a fundamental bottom-up approach for understanding the quantum aspect of gravity. In this paper, we present a framework for the superposition causal diamond spacetime and analyze the behavior of quantum entanglement influenced by the spacetime superposition from the perspective of relativistic quantum information. For the first time,  we combine the concept of spacetime superposition with causal diamonds and derive the analytical expression of the Unruh-diamond vacuum state for Dirac fields in the superposed causal diamond spacetime. Based on this, we analyze both initially correlated and uncorrelated modes in superposed and classical causal diamond spacetimes, and quantify how quantum thermal effects arising from spacetime structure alter entanglement. Our findings reveal that quantum entanglement degrades in classical diamond spacetime, while the superposing structure of spacetime generates additional entanglement resources between modes in superposed diamond spacetimes. From a quantum information perspective, our results suggest that the characteristics of spacetime   superposition can serve as valuable resources for performing quantum information processing tasks.

	\end{abstract}
	
	\pacs{~}
	
	\maketitle

	\section{Introduction}
	
	The compatibility between quantum theory and general relativity constitutes one of the central dilemmas in contemporary physics, as they propose distinct approaches to understand  reality in physics, and their unification might lead to a completely novel understanding of the universe. It has also prompted physicists to develop a gravity theory that aims to reveal the quantum behavior and microstructure of spacetime, known as  quantum gravity theory \cite{hawking1993euclidean,bose2017spin,belenchia2018quantum,christodoulou2019possibility,carlip2008quantum,bronstein2012republication,belenchia2019information}. Several contemporary ``top-down" complete theories, such as string theory \cite{gubser1998gauge,seiberg1999string,witten1995string} and loop quantum gravity \cite{rovelli2008loop,rovelli2015covariant,thiemann2003lectures}, serve as exemplary attempts to demonstrate the self-consistency of these frameworks. Recent studies have endeavored to integrate the foundational principles of quantum theory with those of general relativity, adopting a ``bottom-up" approach to investigate the structural characteristics of spacetime,  to achieve a more profound comprehension of the quantum nature of gravity. This includes quantum features induced by periodically identified Minkowski spacetime superpositions with different characteristic lengths \cite{foo2023quantum}, and non-thermalization phenomena in spatially translational superposition states explored in the framework of quantum field theory in non-inertial reference systems \cite{foo2023superpositions}. These findings deepen our understanding of the quantum properties of gravity and furnish significant theoretical support for the further advancement of quantum gravity theory.
	
	On the other hand,  the theory of relativistic quantum information discloses the profound relationship between gravity and quantum systems, particularly the remarkable impact of the causal structure of spacetime on quantum entanglement between field modes \cite{fuentes2005alice,alsing2006entanglement,an2024quantum,fan2024quantum,wang2016irreversible,camblong2024entanglement,martinetti2003diamond,blencowe2013effective,liu2022gravity,liu2023entanglement,sen2024entanglement,downes2011entangling,liu2023quantum,liu2024optimal}. It was found that  the causal diamond spacetime resulting from the conformal transformation of Rindler spacetime causes a fundamental decoherence of the quantum system due to the presence of an apparent horizon \cite{camblong2024entanglement,martinetti2003diamond}, which affects the efficiency of performing quantum information processing tasks in the spacetime\cite{wilde2013quantum,plenio1998teleportation,horodecki2009quantum}, revealing a far-reaching effect of the spacetime structure on quantum correlations. As a result, we believe that causal diamond spacetime can provide a unique and insightful perspective to studying quantum gravity. On the contrary, one can't detect genuine quantum gravity effects by purely analyzing classical spacetime in the context of relativistic quantum information. To achieve a comprehensive understanding of the quantum nature of gravity emerging from the geometric structure of spacetime, it is essential to apply the principle of superposition from quantum mechanics to the spacetime framework of general relativity.
	
	In this paper, we investigate the generation of entanglement for Dirac fields  in  superposed diamond spacetime. For the first time,  we  integrate the conception of spacetime  superposition with causal diamonds and derive the analytical expression of the Unruh-diamond vacuum state for Dirac fields in the causal diamonds spacetime, thereby establishing a foundational framework for the study of entanglement behavior. We assume that Alice, an inertial observer, and David, a stationary observer with a finite lifetime in causal diamond spacetime, share an entangled state initially. It is demonstrated that quantum entanglement degrades for the Dirac fields in classical diamond spacetime due to the diamond observers limited causal access. Since the diamond observer is in the quantum superposition of the diamond spacetimes'  localized stationary trajectories,   the initial state also undergoes the influence of a superposed quantum channel. In the superimposed causal diamond spacetime, we conduct quantitative analyses of entanglement degradation and generation for initially correlated and uncorrelated modes. Our results demonstrate that the quantum superposition structure of diamond spacetime indeed induces entanglement therefore can improve the efficiency of performing quantum information tasks.
		
		The paper is structured as follows. In Sec. \ref{Sec.2}, we give a conformal transformation between Rindler spacetime coordinates and causal diamond spacetime coordinates. In Sec. \ref{Sec.3}, we analyze the entanglement degradation in the  classical diamond spacetime. In Sec. \ref{Sec.4}, we examine the entanglement of quantum systems in quantum superposition diamond spacetime and classical diamond spacetime for initially correlated and initially uncorrelated modes, respectively. In Sec. \ref{Sec.5}, we presents the conclusions.
		
\section{Coordinates and quantum thermal effects in causal diamond spacetime}\label{Sec.2}
		
\begin{figure}[H]
			\centering
			\includegraphics[scale=0.85]{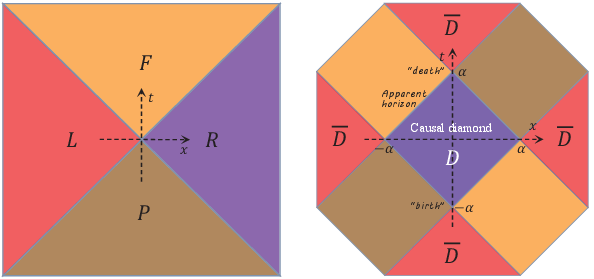}
			\caption{Conformal mapping between Rindler and diamond spacetime regions, the right wedge R $\equiv\{(x_{R};t_{R}):|t_{R}|\leq x_{R}\text{ and }x_{R}\geq0\}$ (in purple) maps into the region D $\equiv\{(x_{D};t_{D}):|t_{D}|+|x_{D}|\leq\alpha $. In the picture on the right, the diamond region D is defined by the intersection of the future light cone of the ``birth" event and the past light cone of the ``death" event, with the apparent horizon as its bounded interface.} \label{RD}
		\end{figure}

The causal diamond's geometry is the overlapping region between the future light cone of the ``birth" event and the past light cone of the ``death" event for the finite lifetime observer \cite{martinetti2003diamond,ida2013modular,su2016spacetime,chakraborty2022thermal}, as shown on the right side of Fig. \ref{RD}. In this spacetime area, the lifetime of the observer is $\mathcal{T}=2\alpha $, with causal access is restricted within the apparent horizon bounded by the light cones. 
		The fundamental principle of parametrizing diamond geometry is  a one-to-one conformal mapping between the right Rindler wedge R $\equiv\{(x_{R};t_{R}):|t_{R}|\leq x_{R}\text{ and }x_{R}\geq0\}$ and the diamond region D $\equiv\{(x_{D};t_{D}):|t_{D}|+|x_{D}|\leq\alpha $ in the Minkowski spacetime \cite{francesco2012conformal,hislop1982modular}. 
		
		The conformal mapping of the Rindler right wedge R to the causal diamond region D is generally formed by the composite of three modules: the special conformal transformation $K(\rho)$, the scaling transformation $\Lambda(\lambda)$, and the spatial translation $T(\alpha)$, which are combined to construct the one-to-one mapping $(t_{R},x_{R})\longrightarrow(t_{D},x_{D})$ of these two spacetime regions. The mapping consists of the composite $M(a;\lambda)=T(-a)\circ K(\frac{1}{2a})\circ\Lambda(\lambda)$, which yields
		\begin{equation}
			\begin{gathered}
				{t_D}={\lambda}\frac{t_R}{(x_R/\tilde{\alpha}+1)^2-(t_R/\tilde{\alpha})^2},\\	
				{x_{D}}=-{\lambda}\frac{(1/2\tilde{\alpha})(\tilde{\alpha}^2-{x_{R}}^2+{t_{R}}^2)}{({x_{R}}/\tilde{\alpha}+1)^2-({t_{R}}/\tilde{\alpha})^2},\label{eq1}
			\end{gathered}
		\end{equation}
		where  $ \lambda $ is scaling factor, and $\tilde{\alpha}=2\alpha/\lambda $. To define the diamond coordinates, we need to get the inverse transformation by inverting the composite mapping
		\begin{equation}
			\begin{gathered}
				{t_{R}}=\frac1{\lambda}\frac{4t_{D}}{(x_{D}/\alpha-1)^2-(t_{D}/\alpha)^2},\\	{x_{R}}=\frac1{\lambda}\frac{(2/{\alpha})
					({\alpha}^2-{x_{D}}^2+{t_{D}}^2)}{(x_{D}/\alpha-1)^2-(t_{D}/\alpha)^2}.\label{eq2}
			\end{gathered}
		\end{equation}
		The mapping between these two coordinates can then be obtained by using the standard transformation relation between the Minkowski coordinates $({t_{R}},{x_{R}})$ of the wedge region and the Rindler coordinates $(\eta,\xi)$ \cite{birrell1984quantum,takagi1986vacuum,crispino2008unruh}
		\begin{equation}
			{t_R}=\frac1\lambda2\epsilon \alpha e^{2\xi/\alpha}\sinh(2\eta/\alpha),
			\enspace{x_{R}}=\frac1\lambda2\epsilon\alpha e^{2\xi/\alpha}\cosh(2\eta/\alpha),\label{eq3}
		\end{equation}
		where $\epsilon=\pm1$ for D and $\overline{\mathrm{D}}$ respectively, $\eta,\xi\in(-\infty,\infty)$, and $\xi $ = constant represents an uniformly accelerated observer with acceleration $2(\alpha e^{2\xi/\alpha})^{-1}$.  Here $a=\frac{2}{\alpha}$ is the Rindler acceleration. The $(\eta,\xi)$ coordinates system is equally applicable in causal diamond spacetime, since by the conformal mapping, every possible value in $(\eta,\xi)$ has a unique spacetime point in diamond spacetime.
		
		Interchanging the above temporal and spatial coordinates $({t_{R}}\leftrightarrow{x_{R}})$, one obtains the regions $\overline{\overline{\mathrm{D}}}$ of the Rindler wedges F and P conformally mapped with coordinates similar to Eq. \ref{eq3}. Furthermore, this conformal transformations do not affect the causal structure of the diamond spacetime, which is a critical advantage of this method. Also, by employing the conformal transformation, the mapping from the Rindler spacetime to the diamond spacetime is one-to-one, which covers the whole Minkowski spacetime \cite{birrell1984quantum,crispino2008unruh,olson2011entanglement}.
		
		The aforementioned conformal mapping allows for a simplified rewriting of the mapping between coordinates by introducing light-cone variables, and facilitates the field quantization of the diamond spacetime below.\underline{}
		The light-cone coordinates are expressed as
		\begin{equation}
			U_{\sigma}=t+\sigma x,
			\
			\tilde{U}_\sigma={t_{R}}+\sigma{x_{R}},
			\
			u_\sigma=\epsilon(\eta+\sigma\xi),
		\end{equation}
		where $\sigma = \pm1$ denotes the propagation direction, corresponding to the left and right shifts, respectively, and  $\epsilon=\pm1$ makes the diamond spacetime's null coordinates always point to the future. Also, for different values of $\sigma $, it can represent the null coordinates of the Minkowski, Rindler, and diamond spacetimes.
		Then based on these coordinates, the expression of Eq. \ref{eq2} can be restated
		\begin{equation}
			\frac{\tilde{V}}{\tilde{\alpha}}=\frac{1+V/\alpha}{1-V/\alpha} ,
			\quad
			\frac{\tilde{U}}{\tilde{\alpha}}=-\frac{1-U/\alpha}{1+U/\alpha}.\label{eq5}
		\end{equation}
		\noindent
		According to Eq. \ref{eq2} and Eq. \ref{eq5}, we can obtain the corresponding mapping of other Rindler wedges to the coordinates of the remaining region of causal diamond spacetime in Minkowski coordinates, as shown in Fig. \ref{RD}. According to Eq. \ref{eq3}, the light-cone variable $u_{\sigma}$ of the diamond spacetime with the light-cone variable $U_{\sigma}$ of the Minkowski spacetime has the following mapping connection in the diamond's internal region D
		\begin{equation}
			e^{2v/\alpha}=\frac{1+V/\alpha}{1-V/\alpha},
			\quad
			e^{2u/\alpha}=\frac{1+U/\alpha}{1-U/\alpha}.
		\end{equation}
		
	And	for the entangled diamond exterior $\overline{\mathrm{D}}$, one obtains
		\begin{equation}
			e^{2\bar{v}/\alpha}=\frac{V/\alpha-1}{V/\alpha+1},
			\quad
			e^{2\bar{u}/\alpha}=\frac{U/\alpha-1}{U/\alpha+1}.
		\end{equation}
		Finally, the outer regions $\overline{\overline{\mathrm{D}}}$ can also be obtained by analytic continuation from regions D and $\overline{\mathrm{D}}$, which corresponds to the F and P regions of the Rindler spacetime.
		
		To investigate the entanglement nature of  the massless, minimally coupled Dirac fields in causal diamond spacetime, we should solve 
		the Dirac equation \cite{birrell1984quantum} 
		\begin{equation}
			(i \gamma^e \partial_e - m) \psi = 0,
		\end{equation}
		in different coordinates. In this equation, $ \psi $ is the Dirac spinor, $ \gamma^e $ are Dirac matrices, $ m $ is the mass of the particle, and $ \partial_e $ is the partial derivative operator.
		
 Initially, we  utilize the light-cone coordinates of Minkowski spacetime ($U_{\sigma} $) and diamond spacetime ($ u_\sigma$) to obtain the positive-frequency wave modes of these two spacetimes. Subsequently, we can superpose and expand the mode function $ \psi $ of the Dirac fields with the wave modes of the above two spacetimes \cite{birrell1984quantum,takagi1986vacuum,crispino2008unruh}, respectively. 
		Given that the regions $ \mathrm{D} $ and $\overline{\mathrm{D}}$ within the causal diamond spacetime are causally disconnected, we establish a connection between them by using global Minkowski modes through the analytic continuation technique previously applied by Unruh in Rindler spacetime  \cite{unruh1976notes}. Then the Bogoliubov transformation relating the Unruh-diamond modes to the diamond modes can be derived by quantizing the Dirac fields using the obtained Unruh-diamond modes.  After normalizing the state vector, the Unruh-diamond vacuum state is found to be \cite{wald1994quantum}
		\begin{equation}
			|0\rangle^U=\cos r\left|0_D,0_{\overline{D}}\right\rangle+\sin r\left|1_D,1_{\overline{D}}\right\rangle,\label{eq9}
		\end{equation}
		where $\tan r=e^{-\pi\omega\alpha/2}$.
		The first excited state is  described as 
		\begin{equation}  	        |1\rangle^U=|1_D,0_{\overline{D}}\rangle.
		\end{equation}
		
		In the following we utilize italics $(D\mathrm{~and~}\overline{D})$ for modes and Roman characters ($\mathrm{D~and~\overline{D}}$) for regions.
		
		\section{Entangled degradation of the causal diamond spacetime}\label{Sec.3}
		
		We consider the two-mode maximally entangled state of the Dirac fields in causal diamond spacetime
		\begin{equation}
			|\psi\rangle_{AU}=\frac{1}{\sqrt{2}}(|0\rangle_{A}|0\rangle_{U}+|1\rangle_{A}|1\rangle_{U}).\label{eq11}
		\end{equation}
		
		In this two-body system, we suppose that observer Alice maintains inertia, while the causal access of the observer David is limited to the causal diamond spacetime. Moreover, Eq. \ref{eq9} has shown that the Unruh-diamond vacuum state is a two-mode entangled state, thus recalculating the $ 	|\psi\rangle_{AD} $  yields
		\begin{equation}
			\begin{aligned}
				|\psi\rangle_{AD\overline{D}}&=\frac{1}{\sqrt{2}}\big(\cos r|0\rangle_{A}|0\rangle_{D}|0\rangle_{\overline{D}}+\sin r|0\rangle_{A}|1\rangle_{D}|1\rangle_{\overline{D}} \big.\\& \big.\quad\quad\quad+|1\rangle_{A}|1\rangle_{D}|0\rangle_{\overline{D}}\big).
			\end{aligned}
		\end{equation}
		
		At this point, mode $ D $ is mapped to both the inner and outer regions of the diamond spacetime.  The system is further divided into three parts: region A, where the inertial observer Alice resides;  region D, where the diamond observer David resides;  and region $\overline{\mathrm{D}}$, where the virtual observer Anti-David resides. Since the inner region D and the outer region $\overline{\mathrm{D}}$ of the diamond spacetime are causally disconnected,  the entanglement with the virtual observer Anti-David is physically inaccessible. Then it is necessary to trace over mode $ \overline{D} $ and obtain
		\begin{equation}
			\begin{aligned}
				\rho_{AD}=\frac1 2(& \cos r^{2}\left|00\right\rangle\left\langle00\right|+\cos r\left(|00\rangle\langle11|+|11\rangle\langle00|\right)\\&\!\!\!\!+\sin r^{2}\left|01\right\rangle\left\langle01\right|+\left|11\right\rangle\left\langle11\right|).\label{eq13}
			\end{aligned}
		\end{equation}
		
		In this paper, we employ logarithmic negativity to quantify the entanglement between the two-body quantum state. The logarithmic negativity value of zero indicates that the quantum state is separable, means that it lacks entanglement characteristics. The logarithmic negativity is denoted as \cite{vidal2002computable,plenio2005logarithmic}
		\begin{equation}
			N(\rho){=}\log_2\lVert\rho^T\rVert, \label{eq14}
		\end{equation}
		where $\left\|\rho^T\right\|$ is the trace norm of the partial transpose matrix $\rho^{T}$ \cite{reed1972methods}.
		
		In addition, we measure the overall correlations of  two-body systems by the mutual information \cite{adami1997neumann,vedral1997quantifying,groisman2005quantum}
		\begin{equation}
			I(\rho_{AB})=S(\rho_A)+S(\rho_B)-S(\rho_{AB}) ,\label{eq15}
		\end{equation}
		where $S=-\operatorname{Tr}(\rho\ln\rho)$ is the von Neuman entropy of the corresponding matrix.
		\begin{figure}[H]
			\centering
			\includegraphics[scale=0.34]{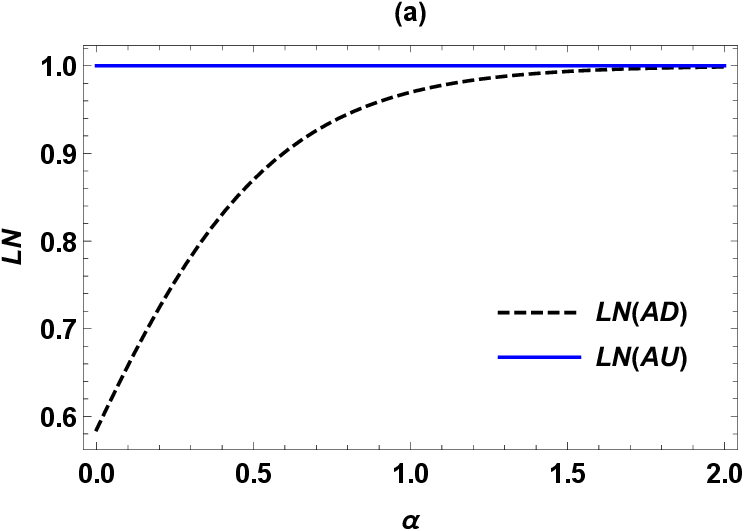}
			\includegraphics[scale=0.34]{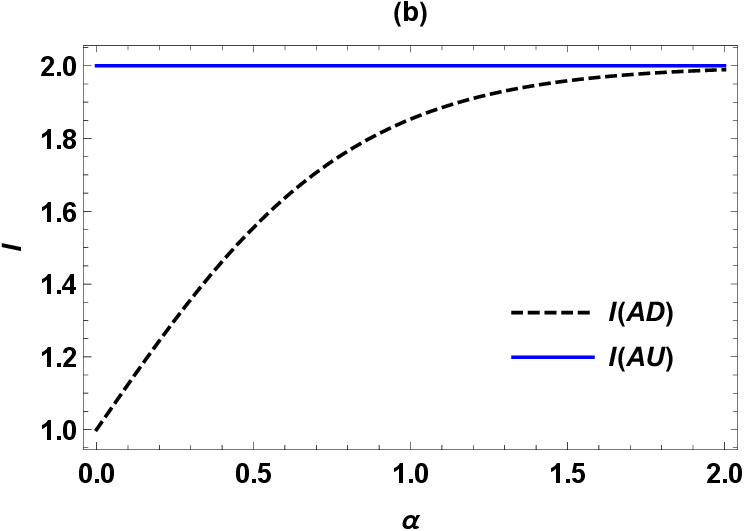}
			\caption{(a): Logarithmic negativity  of quantum states $ \rho_{AU} $ (above) and $ \rho_{AD} $ (below)  as a function of the  parameter $ \alpha $. (b): Mutual information for  quantum states $ \rho_{AU} $ (above) and $ \rho_{AD} $ (below) as a function of the  parameter $ \alpha $.}\label{figBellandD}
		\end{figure}\noindent
		
Fig. \ref{figBellandD} (a) plots the entanglement variation of quantum state $ \rho_{AU} $ and $ \rho_{AD} $ with parameter $ \alpha $. It is demonstrated that for two inertial observers sharing the two-particle state $ \rho_{AU} $, the logarithmic negativity remains invariant irrespective of the parameter $ \alpha $. However, when an inertial observer and a diamond observer share the quantum state $ \rho_{AD} $,  it is observed that the logarithmic negativity serves as an entanglement monotone  \cite{vidal2002computable,plenio2005logarithmic}. Specifically, as $\alpha\to\infty $ (i.e., infinite lifetime, as in Minkowski spacetime), the value is $LN(AD)=1$, which signifies that the two-body subsystem retains maximal entanglement. In contrast, when the parameter $ \alpha $ begins to decrease, the inter-system of entanglement undergoes degeneration. 
		
From the perspective of quantum field theory, it is known that entanglement is observer-dependent. In the diamond spacetime, a finite-lifetime observer is confined to the causal diamond region and cannot access Dirac field modes outside this region, leading to the formation of a thermal state. Additionally, from the relation $ T_{D}=\frac{2}{\pi\mathcal{T}} $,  the observer's lifetime $\alpha$ is inversely proportional to the diamond temperature $T_{D}$. Consequently, as the lifetime parameter $\alpha$ decreases, the temperature $T_{D}$ increases, resulting in the degradation of entanglement within the subsystem. Fig. \ref{figBellandD}(b) analyzes the total correlation between the two subsystems using mutual information, and the findings are consistent with the entanglement analysis mentioned above.
We now know that the thermal effects of diamond spacetime can have an intuitive negative impact on entanglement.  If we represent the change from Eq. \ref{eq11} to Eq. \ref{eq13} as a quantum channel, we may conclude that it is a decay channel.
		
\section{Spacetime superposition produces quantum entanglement}\label{Sec.4}
\subsection{Analysis of entanglement between initially correlated modes}\label{Sec.41}
		
In the previous section, we have shown the entanglement degradation of quantum systems in a single causal diamond spacetime. To find the nature of the spacetime structure, we analyze the behavior of entanglement  in the quantum superimposed diamond spacetime, .
		\begin{figure}[H]
			\centering
			\includegraphics[scale=0.57]{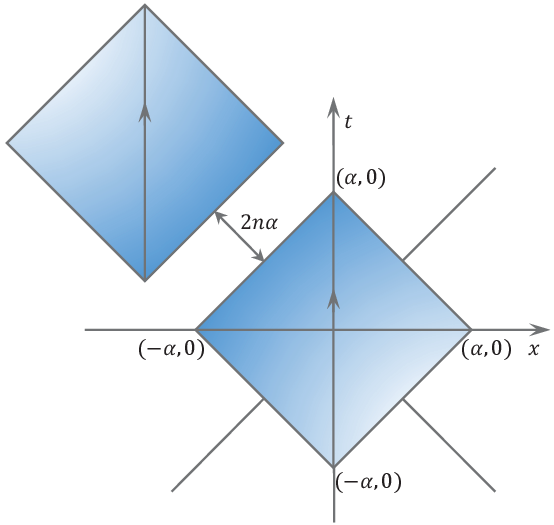}
			\caption{Schematic diagram of the diamond coordinates parameterized by the $ \alpha $.  The zeroth diamond, centered at the origin of Minkowski coordinates, and the $ n $th diamond,  corresponding to a translation of  in the null coordinate are  localized  in the spacetime. And the $ n $th and $ (n+1) $th diamonds share a common boundary.}\label{SD}
		\end{figure}\noindent
		
		We assume  the observer Alice remains inertial, while the diamond observer David is in a quantum superposition of the zeroth and $ n $th diamond localized stationary trajectories, each with the same lifetime value but shifted by a constant offset 2$ n\alpha $. Simultaneous, we introduce a quantum degree of freedom (DoFs) $ f $ to control the movement trajectory followed by the diamond observer David, and the control system is in the superposition case, indicating that the diamond observer follows the superposition trajectory. 
		
		For convenience, we analyze David's trajectory as a superposition of zeroth and oneth diamond spacetime. So far, we define the control system as a quantum superposition by $\left|0\right\rangle $ and $\left|1\right\rangle $ states, and the overall system's initial state is denoted via
		\begin{equation}	  |\Psi\rangle_{AUc}=|\psi\rangle_{AU}\otimes\frac{1}{\sqrt{2}}(|0\rangle+|1\rangle)_{c}.
		\end{equation}
		At this time, the quantum state $|\psi\rangle_{AU}$ shared by the observers Alice and David will undergo an action in the superposed $ r_{1} $ and $ r_{2} $ channel.
		Naturally, after being affected by the overlapping channel, we obtain
		\begin{equation}
			|\Psi\rangle_{AD\overline{D}c}=\frac{1}{\sqrt{2}}(|\psi_{1}\rangle_{AD\overline{D}}|0\rangle_{c}+|\psi_{2}\rangle_{_{AD\overline{D}}}|1\rangle_{c}),\label{eq17}
		\end{equation}
		where
		\begin{equation}
			\begin{aligned}
				|\psi_{1}\rangle_{AD\overline{D}}=\frac{1}{\sqrt{2}}(\cos r_{1}|000\rangle_{AD\overline{D}}+\sin r_{1}|011\rangle_{_{AD\overline{D}}}+|110\rangle_{_{AD\overline{D}}}),
				\\
				|\psi_{2}\rangle_{AD\overline{D}}=\frac{1}{\sqrt{2}}(\cos r_{2}|000\rangle_{AD\overline{D}}+\sin r_{2}|011\rangle_{_{AD\overline{D}}}+|110\rangle_{_{AD\overline{D}}}).
			\end{aligned}
		\end{equation}
		Similarly, tracing over the mode $ \overline{D} $ in the exterior region of diamond , one obtains
		\begin{equation}\begin{aligned}
				\rho_{ADc}&=\frac{1}{2}\Big[\varepsilon_{11}(\rho_{\psi})\otimes|0\rangle_{c}\langle0|+\varepsilon_{12}(\rho_{\psi})\otimes|0\rangle_{c}\langle1|\\&\quad+\varepsilon_{21}(\rho_{\psi})\otimes|1\rangle_{c}\langle0|+\varepsilon_{22}(\rho_{\psi})\otimes|1\rangle_{c}\langle1|\Big],\end{aligned}
		\end{equation}
		where
		\begin{equation}
			\varepsilon_{ij}(\rho_{\psi}):=\mathrm{Tr}_{\overline{D}}\left[|\psi_{i}\rangle_{AD\overline{D}}\langle\psi_{j}|\right],
		\end{equation}
		with $i,j\in\{1,2\}$.\\
		Then, we perform a projection measurement on the control system with the superposition basis $|\pm\rangle_{c}=\frac{1}{\sqrt{2}}(|0\rangle\pm|1\rangle)_{c}$. Once the measurement is finished, the remaining particles will collapse to
		\begin{equation}\begin{aligned}
				\rho_{AD}^{\pm}& = _{c}\langle\pm|\rho_{\psi}|\pm\rangle_{c}/\mathrm{Tr}[ _{c}\langle\pm|\rho_{\psi}|\pm\rangle_{c}] \\
				&=\frac{\varepsilon_{11}(\rho_{\psi})\pm\varepsilon_{12}(\rho_{\psi})\pm\varepsilon_{21}(\rho_{\psi})+\varepsilon_{22}(\rho_{\psi})}{\mathrm{Tr}[\varepsilon_{11}(\rho_{\psi})\pm\varepsilon_{12}(\rho_{\psi})\pm\varepsilon_{21}(\rho_{\psi})+\varepsilon_{22}(\rho_{\psi})]}.\label{eq21}
		\end{aligned}\end{equation}
		Then we find that the quantum states $ \rho_{AD}^{+} $ and $ \rho_{AD}^{-} $ are explicitly denoted as 
		\begin{equation}
			\begin{aligned}
				\rho_{AD}^{+}&=\frac{1}{2M}\Big[\left(\cos r_{1}+\cos r_{2}\right)^{2}|00\rangle_{AD}\langle00|+2\left(\cos r_{1}+\cos r_{2}\right)\\&\quad\quad\quad\otimes(|00\rangle_{AD}\langle11|+|11\rangle_{AD}\langle00|)+\left(\sin r_{1}+\sin r_{2}\right)^{2}\\&\quad\quad\quad\otimes|01\rangle_{AD}\langle01|+4|11\rangle_{AD}\langle11|\Big],
			\end{aligned}
		\end{equation}
		and
		\begin{equation}
			\begin{aligned}
				\rho_{AD}^{-}&=\frac{1}{2N}\Big[\left(\cos r_{1}-\cos r_{2}\right)^{2}|00\rangle_{AD}\langle00|+\left(\sin r_{1}-\sin r_{2}\right)^{2}\\&\quad\quad\quad\otimes|01\rangle_{AD}\langle01|\Big],
			\end{aligned}
		\end{equation}
		where $ \rho_{AD}^{-} $  is a separable state, and
		\begin{equation}
			\begin{aligned}&M=\cos r_{1}\cos r_{2}+\sin r_{1}\sin r_{2}+3,
				\\&N=1-\cos r_{1}\cos  r_{2}-\sin r_{1}\sin r_{2}.\end{aligned}
		\end{equation}
		The probability that measures $ |+\rangle_{c} $, $ |-\rangle_{c} $ are $p_{+}=M/4$ and $p_{-}=N/4$, respectively.
		
\begin{figure}[H]
			\centering
			\includegraphics[scale=0.33]{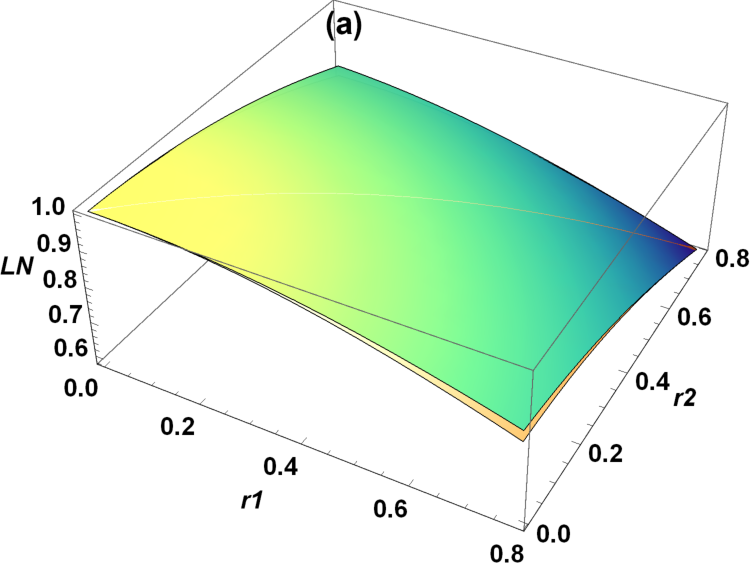}
			\includegraphics[scale=0.33]{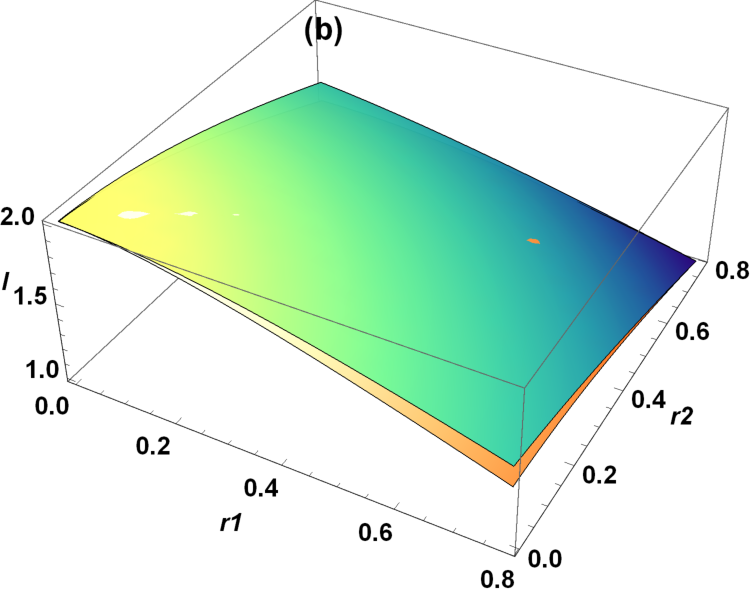}
			\caption{Plots showing (a) average entanglement of logarithmic negativity and (b) average entanglement of mutual information for modes \textit{A} and \textit{D} under the impact of quantum superposition channels (above) and classical hybrid channels (below) varying with parameters $ r_{1} $ and $ r_{2} $. }\label{aenAD}
		\end{figure}

		Employing Eqs. \ref{eq14} and \ref{eq15}, we can calculate the logarithmic negativity $LN(AD)$ and mutual information $I(AD)$ between the initially correlated modes $ A $ and $ D $. Performing multiple measurements on the control system and keeping the measurements, one can obtain the average entanglement $\overline{LN}[{\rho}_{AD}] = p_{+}LN[{\rho}_{AD}^{+}]$ and $\overline{I}[{\rho}_{AD}] = p_{+}I[{\rho}_{AD}^{+}]$. This contrasts with the entanglement from the classical diamond spacetime. When the quantum state $	|\psi\rangle_{AU}$  experiences the action of the classical mixing channel, we obtain
		\begin{equation}
			\begin{aligned}
				\overline{\rho}_{AD}& =\frac{1}{2}\left[\varepsilon_{11}\left(\rho_{AD}\right)+\varepsilon_{22}\left(\rho_{AD}\right)\right] \\
				&=\frac{1}{4}\Big[(\cos r_{1}^{2}+\cos r_{2}^{2})|00\rangle_{AD}\langle00|+\left(\cos r_{1}+\cos r_{2}\right)\\&\quad\quad\otimes(|00\rangle_{AD}\langle11|+|11\rangle_{AD}\langle00|)+(\sin r_{1}^{2}+\sin r_{2}^{2})\\&\quad\quad\otimes|01\rangle_{AD}\langle01|+2|11\rangle_{AD}\langle11|\Big].
			\end{aligned}
		\end{equation}
		Similarly, we can derive the logarithmic negativity $LN[\overline{\rho}_{AD}]$ and mutual information $I[\overline{\rho}_{AD}]$ under the classical hybrid channel. The entanglement varies of quantum superposition channel and classical mixing channel are shown in Fig. \ref{aenAD}.
		\noindent
		
It is shown in Fig. \ref{aenAD} that the average correlations (both classical and quantum correlations) in quantum superposition spacetime are always larger than the average entanglement in classical spacetime between the modes \textit{A} and \textit{D}, regardless of whether we use logarithmic negativity or mutual information. It is commonly known that logarithmic negativity and mutual information between the initially correlated modes are diminishing functions concerning \textit{r}, implying that $ LN(AD) $ and $ I(AD) $ would decrease as \textit{r} increases. It is shown that  the presence of the spacetime structure superposition alleviates entanglement degradation between the initially correlated modes due to quantum  thermal effect of the spacetime, so that $ \overline{LN}[{\rho}_{AD}] \geq LN[\overline{\rho}_{AD}] $ $( \overline{I}[{\rho}_{AD}] \geq I[\overline{\rho}_{AD}])$. This means that the spacetime structure generates additional quantum resources, which is important for performing quantum information processing tasks in spacetime. Furthermore, we discover that  $ \overline{LN}[{\rho}_{AD}]=LN[\overline{\rho}_{AD}] $ ($\overline{I}[{\rho}_{AD}]=I[\overline{\rho}_{AD}] $) at $ r_{1}=r_{2} $, which means that in the limit of $n\to 0 $ and $\omega\rightarrow\omega^{\prime}$, one recovers to a single spacetime. It is worth mentioning that the greater the difference between the parameters $ r_{1} $ and $ r_{2} $, the more noticeable the increase in entanglement and mutual information. 
		
\subsection{Analysis of entanglement between initially uncorrelated modes}\label{Sec.42}
In the previous subsection, we analyzed the entanglement between initially correlated modes in the context of spacetime superposition. However, in relativistic quantum information theory, examining the behavior of initially uncorrelated modes can provide  a more intuitive understanding of how spacetime structure influences entanglement generation. 		
		
		By tracing over the mode $ A $ from Eq. \ref{eq17}, we  obtain the correlation state between modes $ D $ and  $ \overline{D} $
		\begin{equation}
			\begin{aligned}
				\rho_{D\overline{D}c}&=\frac{1}{2}\Big[\varepsilon_{11}(\rho_{\psi})\otimes|0\rangle_{c}\langle0|+\varepsilon_{12}(\rho_{\psi})\otimes|0\rangle_{c}\langle1|\\&\quad+\varepsilon_{21}(\rho_{\psi})\otimes|1\rangle_{c}\langle0|+\varepsilon_{22}(\rho_{\psi})\otimes|1\rangle_{c}\langle1|\Big],
			\end{aligned}
		\end{equation}
		where
		\begin{equation}
			\varepsilon_{ij}(\rho_{\psi}):=\mathrm{Tr}_{A}\left[|\psi_{i}\rangle_{AD\overline{D}}\langle\psi_{j}|\right],
		\end{equation}
		with $i,j\in\{1,2\}$.\\
		Likewise, after completing projection measurements on the control system, we obtain the quantum states $\rho_{D\overline{D}}^{+}$ and $\rho_{D\overline{D}}^{-}$ 		\begin{equation}
			\begin{aligned}
				\rho_{D\overline{D}}^{+}&=\frac{1}{2M}\Big[\left(\cos r_{1} +\cos r_{2}\right)^{2}|00\rangle_{D\overline{D}}\langle00|+\left(\cos r_{1} \sin r_{1}\right.\\&\left.\quad\quad\quad+\cos r_{1} \sin r_{2}+\cos r_{2} \sin r_{1}+\cos r_{2} \sin r_{2}\right)\\&\quad\quad\quad\otimes(|00\rangle_{D\overline{D}}\langle11|+|11\rangle_{D\overline{D}}\langle00|)+\left(\sin r_{1}+\sin r_{2}\right)^{2}\\&\quad\quad\quad\otimes|01\rangle_{D\overline{D}}\langle01|+4|10\rangle_{D\overline{D}}\langle10|\Big],
			\end{aligned}
		\end{equation}
		and
		\begin{equation}
			\begin{aligned}
				\rho_{D\overline{D}}^{-}&=\frac{1}{2N}\Big[\left(\cos r_{1} -\cos r_{2}\right)^{2}|00\rangle_{D\overline{D}}\langle00|+\left(\cos r_{1} \sin r_{1}\right.\\&\left.\quad\quad\quad-\cos r_{1} \sin r_{2}-\cos r_{2} \sin r_{1}+\cos r_{2} \sin r_{2}\right)\\&\quad\quad\quad\otimes(|00\rangle_{D\overline{D}}\langle11|+|11\rangle_{D\overline{D}}\langle00|)+\left(\sin r_{1}-\sin r_{2}\right)^{2}\\&\quad\quad\quad\otimes|11\rangle_{D\overline{D}}\langle11|\Big].
			\end{aligned}
		\end{equation}
		The probability measured in $ |+\rangle_{c} $, $ |-\rangle_{c} $ between the initial uncorrelated modes are  $p_{+}=M/4$ and $p_{-}=N/4$, respectively. And when the joint state of modes \textit{D} and $ \overline{D} $ undergoes the action of the classical hybrid channel, one obtains
		\begin{equation}
			\begin{aligned}
				\overline{\rho}_{D\overline{D}}& =\frac{1}{2}\left[\varepsilon_{11}(\rho_{D\overline{D}})+\varepsilon_{22}(\rho_{D\overline{D}})\right] \\&=\frac{1}{4}\Big[(\cos r_{1}^{2}+\cos r_{2}^{2})|00\rangle_{D\overline{D}}\langle00|+\left(\cos r_{1}\sin r_{1}\right.\\&\left.\quad\quad+\cos r_{2}\sin r_{2}\right)(|00\rangle_{D\overline{D}}\langle11|+|11\rangle_{D\overline{D}}\langle00|)\\&\quad\quad+(\sin r_{1}^{2}+\sin r_{2}^{2})|11\rangle_{D\overline{D}}\langle11|+2|10\rangle_{D\overline{D}}\langle10|\Big].
			\end{aligned}
		\end{equation}
		
		By combining the quantum states acquired above with Eqs. \ref{eq14} and \ref{eq15}, we can calculate the average entanglement $\overline{LN}[{\rho}_{D\overline{D}}] = p_{+}LN[{\rho}_{D\overline{D}}^{+}]+p_{-}LN[{\rho}_{D\overline{D}}^{-}]$ and $\overline{I}[{\rho}_{D\overline{D}}] = p_{+}I[{\rho}_{D\overline{D}}^{+}]+p_{-}I[{\rho}_{D\overline{D}}^{-}]$ under the quantum superposition channel, as well as the logarithmic negativity $ LN[\overline{\rho}_{D\overline{D}}] $ and mutual information $ I[\overline{\rho}_{D\overline{D}}] $ for the classical hybrid channel for the initial uncorrelated modes. 
		\begin{figure}[H]
			\centering
			\includegraphics[scale=0.33]{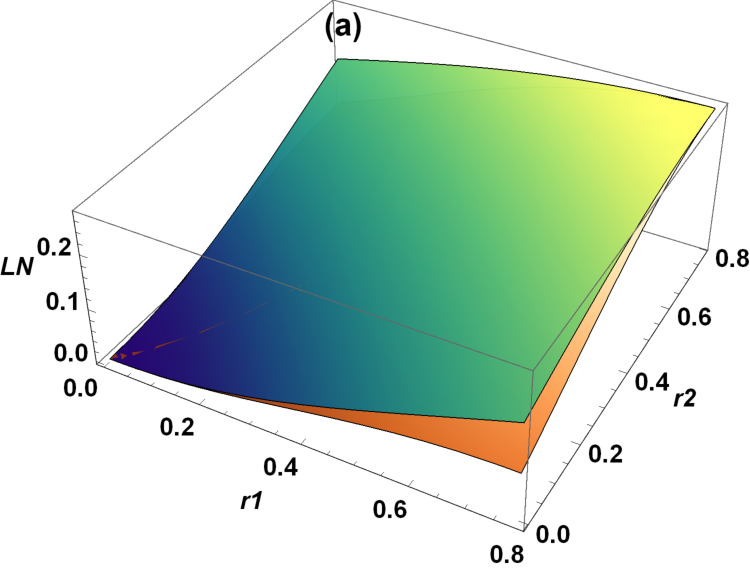}
			\includegraphics[scale=0.33]{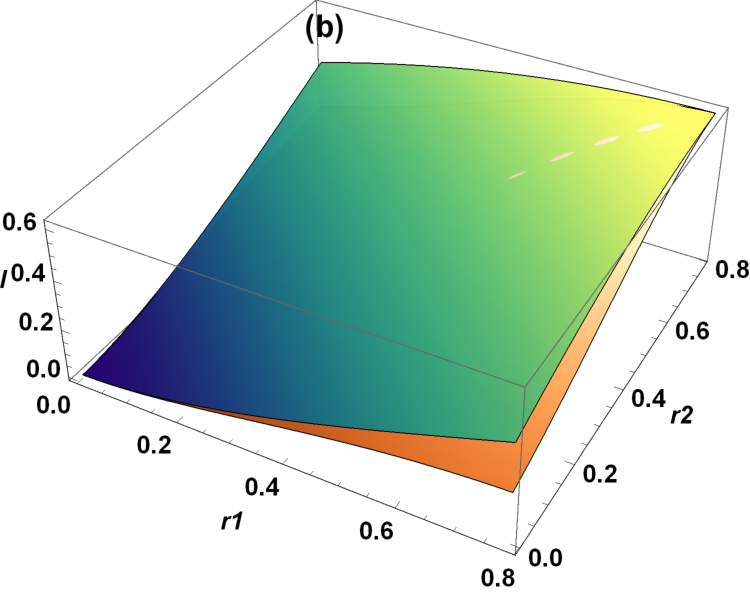}
			\caption{Plots showing (a) $ \overline{LN}[{\rho}_{D\overline{D}}]$ (above), $LN[\overline{\rho}_{D\overline{D}}] $ (below) and (b)  $ \overline{I}[{\rho}_{D\overline{D}}]$ (above), $ I[\overline{\rho}_{D\overline{D}}] $ (below) for the joint states of uncorrelated modes $ D $ and $ \overline{D} $ varying with parameters $ r_{1} $ and $ r_{2} $.}\label{aenDtildeD}
		\end{figure}\noindent
		
		In Fig \ref{aenDtildeD}, we find that when analyzing the joint states of initially uncorrelated modes \textit{D} and $\overline{D}$ under both the quantum superposition channel and the classical hybrid channel, the entanglement in quantum superposition spacetime is consistently greater than that in classical spacetime, i.e. $ \overline{LN}[{\rho}_{D\overline{D}}] \geq LN[\overline{\rho}_{D\overline{D}}] $ $( \overline{I}[{\rho}_{D\overline{D}}] \geq I[\overline{\rho}_{D\overline{D}}]$).  It is well established that increasing the value of the parameter $r$ can induce entanglement between initially uncorrelated modes. Upon quantifying the conditions for entanglement generation, it becomes evident that the spacetime structure in quantum superposition spacetime provides additional resources for quantum entanglement compared to classical spacetime.
		
		\section{Conclusion}\label{Sec.5}
		
		In this paper, we investigate the dynamics of entanglement in both superposed and classical causal diamond spacetimes for massless Dirac fields. Our findings reveal that the finite-lifetime-induced thermal effect experienced by an observer in classical diamond spacetime leads to entanglement degradation, thereby diminishing the performance of quantum information processing tasks. It is shown that quantum entanglement in superposed diamond spacetime exceeds that in classical diamond spacetime, which indicates that the superposition structure of diamond spacetime generates additional entanglement resources, mitigates thermal-induced entanglement degradation, and enhances the efficiency of quantum information processing tasks. The analysis presented herein offers a bottom-up perspective on the quantum properties arising from the superposition structure of spacetime, providing evidence that spacetime's nature may be inherently quantum. This contributes to the unification of general relativity and quantum mechanics, thus holding significant theoretical implications.

		\acknowledgments
		
		This work was supported by the National Natural Science Foundation of China under Grants No.12475051, No.12374408, and No.12035005; the science and technology innovation Program of Hunan Province under grant No.2024RC1050; the Natural Science Foundation of Hunan Province under grant No.2023JJ30384; and the innovative research group of Hunan Province under Grant No.2024JJ1006.

		\bibliographystyle{apsrev4-1}
		%

	\end{document}